\documentstyle[aps,epsfig,multicol]{revtex}

\begin{document}
\draft
\title{Ferromagnetism in the Strong Hybridization 
Regime of the Periodic Anderson Model}
\author{C. D. Batista,$^1$ J. Bon\v ca,$^2$ and J. E. Gubernatis$^1$}
\address{$^1$Center for Nonlinear Studies and Theoretical Division, 
Los Alamos National Laboratory, Los Alamos, NM 87545\\
$^2$ Department of Physics, FMF, University of Ljubljana and J.
Stefan Institute, Ljubljana, Slovenia}
\date{\today}
\maketitle
\begin{abstract}

We determine exactly the ground state of the one-dimensional periodic Anderson model 
(PAM) in the strong hybridization regime. In this regime, the low 
energy sector of the PAM maps into an effective Hamiltonian
that has a ferromagnetic ground state for any electron density 
between half and three quarters filling. This rigorous result proves 
the existence of a new magnetic state that was excluded in the previous 
analysis of the mixed valence systems.

\end{abstract}
\pacs{}

\begin{multicols}{2}

\columnseprule 0pt

\narrowtext
\vspace*{-0.5cm}

Rigorous results, numerical and analytic, have greatly aided the study
of strongly correlated electrons systems. Unfortunately, few such
results exist. The numerical renormalization group and Bethe ansatz
solutions of the single impurity Anderson and Kondo models are perhaps
the best examples of a solid numerical approach and exact solution of
simple models that changed and solidified the thinking in what was a
highly controversial and puzzling problem area. Another important result was
the rigorous connection between the two models established by 
Schrieffer and Wolff \cite{Schrieffer}.  Their work showed that the 
low lying energy spectrum of the Anderson model in the strong coupling 
and weak hybridization limit ($U/t\gg 1$ and $V \ll |E_F-\epsilon_f|$) 
can be mapped into the Kondo model in the weak coupling regime ($J/t \ll 1$). 

For dense systems, the natural extensions of the impurity models are
the periodic Anderson (PAM) and Kondo lattice (KLM) models. For these,
numerical renormalization group and Bethe Ansatz solutions are lacking even
in one-dimension. In addition, very few rigorous results are available for the PAM 
\cite{Ueda,Yana}. What remains true however is the connection between the
strong and weak coupling limits via a natural extension of the
Schrieffer-Wolff transformation. 



There are two basic questions one can ask about these lattice models:
what is their relevance to real materials and in what other parameter
regimes might their physics be connected? In several well known
papers, Doniach \cite{Doniach}, at least implicitly, made several assumptions about
the answers to both questions and proposed the now standard picture of
the magnetic properties of f-electron materials that
portrays a competition between the RKKY magnetic interaction which is 
obtained from a fourth order expansion in the hybridization and the Kondo exchange.



The reasoning behind this intuitively appealing picture is something
like the following: From the Schrieffer-Wolff perturbation theory, the
Kondo exchange coupling is related to the parameters in the PAM via $J
\approx |V|^2/|E_F-\epsilon_f|$. In the PAM, a mixed valence regime
corresponds to positioning the f-electron orbitals in the conduction band
near the Fermi energy, i.e., $|E_F-\epsilon_f| \approx 0$. In this
regime, the Schrieffer-Wolff result suggests the Kondo exchange is
strong and thus can lead to a complete compensation of the f-moments
by one or more of the conduction band electrons. Implied in this line
of reasoning there are two important assumptions \cite{Doniach}: I) 
The strong coupling limit 
of the KLM  is connected to the mixed valence regime 
(for weak hybridzation $|V| \ll |t|$) of the PAM, and II) the number 
of conduction electrons per $f$-magnetic moment is larger than one.
Regarding the first assumption, we note that this connection 
is not established by the Schrieffer-Wolff transformation because 
when $|E_F-\epsilon_f| \approx 0$, the transformation is no longer valid.


We also note recent numerical studies \cite{nos1,nos2} of the weak hybridzation, strongly
coupled PAM that find over a wide range of parameters ferromagnetic
states which are not a result of the RKKY interaction and non-magnetic
states which are not a consequence of a Kondo-like compensation of the
f-moments by the conduction band electrons \cite{Bonca}. Additionally, there are
mixed valence materials which exhibit a co-existence of ferromagnetism
and a strong Kondo-like behavior (e.g., CeSi$_{x}$ \cite{Yashima}, CeGe$_2$ \cite{Lahiouel},
Ce(Rh$_{1-x}$Ru$_x$)$_3$B$_2$ \cite{Malik}, CeSi$_{1.76}$Cu$_{0.24}$ \cite{Boni}, 
Ce$_3$Bi$_4$, and CeNi$_{0.8}$Pt$_{0.2}$\cite{Fillion}). Because the RKKY interaction is
considerably weaker than the Kondo exchange, these experimental
results suggest that a ferromagnetic mechanism other
than RKKY, at least sometimes, dominates in mixed valence materials.


In this paper we note that a strong hybrization, $|V|\gg |t|$, can lead
to mixed valence state and a one-on-one compensation of an f-moment by
a conduction electron. Then, for the strong couping limit of the 
one-dimensional KLM we rigorously establish that the ground state is
connected to the strong hybridzation, strong coupling limit of the
PAM. We thus show that Doniach's first
assumption is valid if the hybridization is strong, although its validity 
remains questionable for the more relevant regime $|V|\ll |t|$ (weak hiybridization).
In the strong hybridization limit we 
rigorously show that there is  ferromagnetic ordering instead of a 
nonmagnetic Kondo state. The difference between our result and
Doniach's can be attributed to the violation of his second assumption
which indeed is not valid for a lattice system (one $f$-orbital per unit cell) 
since the number of conduction electrons (with net magnetic moment) 
per $f$ spin  cannot be larger than one (it is equal to one only 
at half-filling).


The Hamiltonian for the one dimensional PAM is:
\begin{eqnarray}
H &=& -t \sum^{L-1}_{{\bf i}=1,\sigma} (d_{{\bf i}\sigma}^\dagger
  d^{}_{{\bf i}+1\sigma}+ d_{{\bf i}+1\sigma}^\dagger d^{}_{{\bf i}\sigma})
  +  \epsilon_f \sum^{L}_{{\bf i}=1,\sigma} n_{{\bf i}\sigma}^f 
\nonumber \\
  &-& V \sum^{L}_{{\bf i}=1,\sigma} (d_{{\bf i}\sigma}^\dagger
  f^{}_{{\bf i}\sigma}+ f_{{\bf i}\sigma}^\dagger
  d^{}_{{\bf i}\sigma})+ \frac{U}{2}
  \sum^{L}_{{\bf i}=1,\sigma}n_{{\bf i}\sigma}^f n_{{\bf i}\bar {\sigma}}^f\ ,
\nonumber
\end{eqnarray}
where $d_{{\bf i}\sigma}^\dagger$ and $f_{{\bf r}\sigma}^\dagger$ create an
electron with spin $\sigma$ in the $d$ and $f$ orbitals of the lattice
site ${\bf i}$ and $n^f_{{\bf i}\sigma}=f^{\dagger}_{{\bf i}\sigma}f^{\;}_{{\bf i}\sigma}$.

We will show that for infinite $U$, $|t| \ll |V|$, and 
$|\epsilon_f| \lesssim |V| $ (asymmetric regime), the 
low energy sector of the PAM can be mapped into an infinite
$U$ Hubbard model which includes correlated next-nearest 
neighbor hoppings and nearest-neighbor 
repulsions. To this end we first need to solve the atomic 
(one site) limit of the PAM \cite{Chan} for all the possible fillings 
(0 to 3 particles per site because $U$ is infinite).

\begin{figure}[tbp]
\begin{center}
\vspace{-0.1cm} 
\epsfig{file=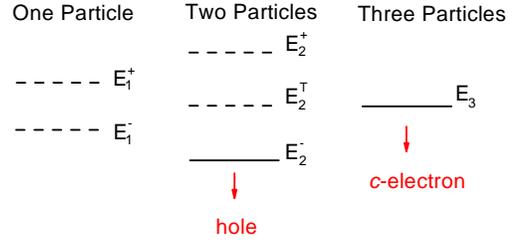,height=75mm,angle=-90}
\end{center}
\vspace{-2.3cm} \caption{Energy levels for the atomic solution of the PAM for 
infinite $U$. The full lines correspond to the lowest energy levels 
that generate the reduced subspace of $H_{eff}$.} 
\label{fig1}
\end{figure}

There are two possible eigenstates for one particle on one site ${\bf i}$. 
These eigenstates are created by the following operators:
\begin{equation}
 \alpha_{{\bf i}\sigma}^\dagger
    =  u f_{{\bf k}\sigma}^\dagger
                     + v d_{{\bf i}\sigma}^\dagger, \; \; \;
 \beta_{{\bf i}\sigma}^\dagger
    =  - v f_{{\bf i}\sigma}^\dagger
                     + u  d_{{\bf i}\sigma}^\dagger,
\label{alphak}
\end{equation}
with
\begin{equation}
u= \frac{E_1^{+} - \epsilon_f}{\sqrt{(E_1^{+} - \epsilon_f)^2+V^2}}, \; \;
v= \frac{V} {\sqrt{(E_1^{+} - \epsilon_f)^2+V^2}}.
\\ \nonumber
\label{uv}
\end{equation}
The operator $\alpha_{{\bf i}\sigma}^\dagger$ creates a particle in the 
bonding state with energy 
$E_1^{-}=\frac{\epsilon_f}{2}-\sqrt{\frac{\epsilon_f^2}{4}+V^2}$,
while $\beta_{{\bf i}\sigma}^\dagger$ creates a particle in the 
anti-bonding state with energy $E_1^{+}=\frac{\epsilon_f}{2}+\sqrt{\frac{\epsilon_f^2}{4}+V^2}$
(see Fig.~\ref{fig1}). 

For two particles on one site, there are two singlets ($U=\infty$), and three triplet eigenstates 
(see Fig.~\ref{fig1}). The ground and the highest energy states are the bonding and
the anti-bonding singlets respectively: 
\begin{eqnarray}
|\phi^{-}\rangle &=& \frac {a}{\sqrt{2}} 
(f^{\dagger}_{{\bf i}\uparrow}d^{\dagger}_{{\bf i}\downarrow}
- f^{\dagger}_{{\bf i}\downarrow} d^{\dagger}_{{\bf i}\uparrow}) 
+ b \; d^{\dagger}_{{\bf i}\uparrow} d^{\dagger}_{{\bf i}\downarrow},
\\ \nonumber 
|\phi^{+}\rangle &=& -\frac {b}{\sqrt{2}} 
(f^{\dagger}_{{\bf i}\uparrow}d^{\dagger}_{{\bf i}\downarrow}
- f^{\dagger}_{{\bf i}\downarrow} d^{\dagger}_{{\bf i}\uparrow}) 
+ a \; d^{\dagger}_{{\bf i}\uparrow} d^{\dagger}_{{\bf i}\downarrow}.
\end{eqnarray}
with
\begin{equation}
a= \frac{E_2^{+} - \epsilon_f}{\sqrt{(E_2^{+} - \epsilon_f)^2+2V^2}}, \; \;
b= \frac{\sqrt{2}V} {\sqrt{(E_2^{+} - \epsilon_f)^2+2V^2}},
\\ \nonumber
\label{ab}
\end{equation}
The energies of the bonding and the anti-bonding singlets are: 
$E^{\pm}_2= \frac{\epsilon_f}{2} \pm \sqrt{\frac{\epsilon_f^2}{4}+ 2 V^2}$.
The triplet states:
\begin{eqnarray}
|\phi^{T}_1\rangle= f^{\dagger}_{{\bf i}\uparrow} d^{\dagger}_{{\bf i}\uparrow},
\;\;\;\;\;
|\phi^{T}_{-1}\rangle= f^{\dagger}_{{\bf i}\downarrow} d^{\dagger}_{{\bf i}\downarrow}
\\ \nonumber
|\phi^{T}_0\rangle= \frac {1}{\sqrt{2}}
(f^{\dagger}_{{\bf i}\uparrow} d^{\dagger}_{{\bf i}\downarrow}
+f^{\dagger}_{{\bf i}\downarrow} d^{\dagger}_{{\bf i}\uparrow})
\end{eqnarray}
have energy $E^{T}_2= \epsilon_f$.

For three particles on one site, there are two possible states due to
the two possible orientations of the $f$ spin: 
$f^{\dagger}_{{\bf i}\sigma}d^{\dagger}_{{\bf i}\uparrow}d^{\dagger}_{{\bf i}\downarrow}
|0\rangle$.
The energy of both states is $E_3=\epsilon_f$.

We will consider the range of 
concentrations: $\frac{1}{2} \leq n \leq \frac{3}{4}$, where 
$n=N_e/4L$ and $N_e$ is the 
total number of electrons. In this way, the concentration ranges 
from two to three particles per site.
For $t=0$, the ground state of $H$ is massively 
degenerate, and the corresponding subspace is generated by states 
containing $L(3-4n)$ local bonding singlets $|\phi^{-}\rangle$ (doubly occupied sites)
and $4L(n-\frac{1}{2})$ sites occupied by three particles. Therefore, 
the effective Hamiltonian which is obtained when $t$ is included perturbatively 
has a local dimension equal to 3 because each site can be occupied by a bonding singlet
(hole) or by three particles state with two possible spin orientations (S=1/2 particle). 
The huge degeneracy of the ground state is then associated with both the spin and the 
charge degrees of freedom. We will see below that while the 
degeneracy associated with the charge degrees of freedom is lifted 
to first order in $t$, the spin degeneracy is lifted at second order. 

\begin{figure}[tbp]
\begin{center}
\vspace{-0.1cm} 
\epsfig{file=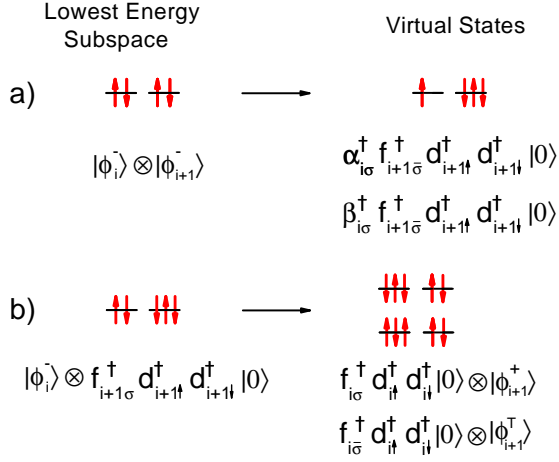,height=80mm,angle=-90}
\end{center}
\vspace{0.0cm} \caption{Virtual states which are obtained from the
application of the hopping term $t$ to the lowest energy 
subspace of $H(t=0)$.} 
\label{fig2}
\end{figure}

To derive an effective Hamiltonian, we first need to identify the 
possible virtual processes.   
There are two different types of virtual (excited) states which are obtained 
by applying the hopping term to the ground state subspace of $H(t=0)$ (see Fig. \ref{fig2}): 
a) those in which two local 
nearest neighbor bonding singlets ($2+2$) are excited to local states having one and three 
particles ($1+3$) and b) those in which a local bonding singlet and a three particle state,
which are nearest neighbors, are permuted and excited to the other possible local states (anti-bonding
singlet or triplet). In any of these virtual processes, $t$ is much smaller than the energy
difference between the virtual and the ground state because $|t| \ll |V|$ and 
$ |\epsilon_f| \lesssim |V| $.  

Introducing the creation and annihilation operators: 
\begin{eqnarray}
c^{\dagger}_{{\bf i} \sigma} &=& (1-n^{f}_{{\bf i}{\bar \sigma} }) f^{\dagger}_{{\bf i} \sigma}
d^{\dagger}_{{\bf i}\uparrow}d^{\dagger}_{{\bf i}\downarrow}, 
\\ \nonumber
c^{\;}_{{\bf i} \sigma} &=& d^{\;}_{{\bf i}\downarrow}d^{\;}_{{\bf i}\uparrow}
f^{\;}_{{\bf i} \sigma}(1-n^{f}_{{\bf i} {\bar \sigma}}), 
\end{eqnarray}
and mapping the bonding singlet, $|\phi^{-}\rangle$, into the empty state or hole,
we can write the effective Hamiltonian up to second order in $t$:
\begin{eqnarray}
H_{eff} &=& - {\tilde t} \sum_{{\bf i},\sigma}
(c^{\dagger}_{{\bf i}+1 \sigma} c^{\;}_{i \sigma} + H.c.)
+ e_0 \sum_{{\bf i}} (1-n_{\bf i})
\\ \nonumber 
&+& t'_1 \sum_{{\bf i},\sigma} 
(c^{\dagger}_{i+1 \sigma} n_i c^{\;}_{i-1 \sigma} + H.c.)
\\ \nonumber
&+& t'_2 \sum_{{\bf i},\sigma,\sigma'} 
(c^{\dagger}_{i+1 \sigma'} c^{\dagger}_{i \sigma} c^{\;}_{i \sigma'} 
c^{\;}_{i-1 \sigma} + H.c.)
\\ \nonumber
&+& t'_3 \sum_{{\bf i},\sigma} 
(c^{\dagger}_{i+1 \sigma} (1-n_{\bf i}) 
c^{\;}_{i-1 \sigma} + H.c.) 
\\ \nonumber
&+& g \sum_{{\bf i},\sigma} (1-n_{\bf i+1})(1-n_{\bf i}) + \epsilon_f L
\label{heff}
\end{eqnarray}
with 
\begin{eqnarray}
{\tilde t} &=& \frac{ta^2}{2},
\;\;\;\;\;\;
t'_1 = \frac{t^2a^2}{2(E^{T}_2-E^{-}_2)},
\;\;\;\;\;\;
\\ \nonumber
t'_2 &=&  \frac {t^2 a^2}{4}
\Biggl [\frac{b^2}{E^{+}_2-E^{-}_2} - \frac{1}{E^{T}_2-E^{-}_2}  \Biggr],
\\ \nonumber
t'_3 &=& \frac{t{\tilde t}}{2} \Biggl [ \frac{(au+\sqrt{2}vb)^2}{\epsilon_f+E^{-}_1-2E^{-}_2}
+ \frac{(av-\sqrt{2}ub)^2}{\epsilon_f+E^{+}_1-2E^{-}_2} \Biggr]
\\ \nonumber
e_0 &=& E^{-}_2 - \epsilon_f - 2 {\tilde e}_0
\;\;\;\;\;\;
g = 2 {\tilde e}_0 - 4 t'_3
\\ \nonumber
{\tilde e}_0 &=& \frac{t {\tilde t}}{2} 
\Biggl [\frac{b^2}{E^{+}_2-E^{-}_2} + \frac{3}{E^{T}_2-E^{-}_2}  \Biggr]
\end{eqnarray}
The half-filled PAM corresponds to the absence of $c$ electrons (vacuum) and
$n=\frac{3}{4}$ corresponds to one $c$ electron per site.

If we only keep the first order terms of $H_{eff}$, the 
resulting Hamiltonian is an infinite $U$ Hubbard model.
In this model, the charge and the spin degrees of freedom 
are decoupled and therefore the wave function is a product  
of a charge and a spin component. As a consequence, 
the infinite $U$ Hubbard Hamiltonian maps into a 
spinless model for each fixed spin configuration. Therefore,
the complete spin degeneracy persists to first order in $t$ 
while the charge degeneracy is lifted.

The second order terms introduce  next nearest neighbor correlated 
hoppings $t'_1$, $t'_2$, $t'_3$  and a repulsion $g$ between
nearest neighbors. 
The terms $t'_1$ and $t'_2$ describe hopping processes where the
electron hops over another electron in the intermediate site with
and without spin flip. The $t'_3$ term describes the hopping
over an empty site. Notice that $t'_1$ is the only term which 
can change the spin ordering.

Using a second order expansion around the strong coupling limit (large
$J/t$) of the Kondo lattice model (KLM), Sigrist et al \cite{Sigrist}
found an effective Hamiltonian for the KLM of the same form as (7).
The present result is more general. For specific choices of PAM
parameters, (7) reduces to their result (and hence the eigenvalue
spectra of two Hamiltonians become identical). The main point is that
the equivalence in forms establishes a formal connection 
between the strong hybridization (mixed valence) and strong coupling regime of the 
PAM, and the strong coupling limit of the KLM. Given this connection, we 
recognize that the bonding singlet, represented by a hole 
in $H_{eff}$, is a Kondo singlet $|\phi^{-}\rangle$ in a mixed valence regime.
It is interesting to remark that the large $J/t$ limit of the KLM has instead been  
associated to the small $U$ limit (weak coupling) of the PAM \cite{Tsunetsugu}. 

Sigrist et al \cite{Sigrist} also proved that to 
leading order in $t'_i$ ($1\leq i \leq3$), the ground state
of $H_{eff}$ is ferromagnetic for any concentration of $c$ electrons.
To this end, they noticed that $t'_1$ is the only relevant term 
since it can change the spin configuration. The particular sign of 
$t'_1$ allowed them to apply the Perron-Frobenius theorem and demonstrate
the ferromagnetic character of the ground state.
Following exactly the same procedure, we can prove that the spin 
degeneracy of the 1D PAM for $t=0$ and $|\epsilon_f| \lesssim|V|$ is 
lifted in a perturbative sense toward a ferromagnetic state for 
any concentration $\frac{1}{2} \leq n \leq \frac{3}{4}$
with a total spin per site $2(n-\frac{1}{2})$. 

The microscopic mechanism for the stabilization of the FM 
ground state is similar to the one found by Nagaoka \cite{Nagaoka}
in the Hubbard model for dimension larger than one. Even though
$H_{eff}$ is a one dimensional model, there are two different 
ways to move one hole to a next-nearest neighbor site by hopping 
over one $c$-electron: either by 
two applications of ${\tilde t}$ or by one application of $t'_1$. Only when
the background is FM will both processes give rise to the same final 
state. If the sign of the final state is the same
for both processes, the effective matrix element is reinforced (constructive
interference) and kinetic energy of the hole is lowered. This is 
indeed what happens in $H_{eff}$ due to the positive sign of $t'_1$.
Therefore, the coherent propagation of the Kondo singlet (hole)
is responsible for the polarization of the spins which are not 
quenched by the conduction electrons. Since this coherent 
propagation is only possible for dense systems where the $f$-orbitals
form a lattice, the stabilization of the FM state is excluded from any 
analysis which only considers the dilute limit (few $f$-orbitals).

The same perturbative analysis used to prove the
FM character of the ground state can also be used to show \cite{Sigrist,Shiba}
that the low energy spin excitations of $H_{eff}$ are described by a FM 
Heisenberg model with an effective exchange interaction: 
${\bar J}=2\frac{t'_1}{\pi}[\frac{2}{\pi\rho}sin^2(\pi\rho)-sin(2\pi\rho)]$,
where $\rho$ is the density of $c$-electrons. This effective Heisenberg
model operates in a truncated Hilbert space of $H_{eff}$ that 
only contains spin degrees of freedom. The separation between 
charge and spin degrees of freedom occurs because
to leading order in $t'_i$, the eigenstates are still factorisable into 
their orbital and spin components. 

In summary, we proved that  the PAM has a ferromagnetic 
ground state in the strong hybridization and strong coupling regime
for concentrations between half and three quarters filling. This wide
region of concentrations clearly indicates that the microscopic mechanism 
is not related to an RKKY interaction as is expected for a strong mixed 
valence regime. Instead, the stabilization of the FM ground state 
is due to the coherent propagation of the Kondo singlets in a FM background.
The mechanism is similar to the one operating in the Nagaoka \cite{Nagaoka} solution of
the infinite $U$ Hubbard model. 

It is important to ask what can we expect in higher dimensions.
In this case, the spin degeneracy of $H(t=0)$ is lifted to first order in $t$
because there is no separation between the charge and the spin degrees of freedom. 
It is well known that the ferromagnetic Nagaoka \cite{Nagaoka} state is stabilized 
when one electron is added to the half filled system. Different works 
\cite{Zhang,Putikka,Liang} indicate that a partially polarized FM state is 
favored for a finite range of concentrations away from half filling. 
In addition, we have recently found numerical evidence of itinerant ferromagnetism
in the PAM for the mixed valence regime  and $|V|\lesssim |t|$ \cite{nos1,nos2}.

The current situation clearly merits reasking what are the possible 
scenarios when a dense system is away from half-filling to evolve
from the localized to the mixed valence regime. 
In the localized regime, the hybridization can 
be considered as a perturbation and the corresponding perturbation theory 
tells us that the magnetic properties are dominated by the RKKY interaction.
When the system approaches the mixed valence regime, there are at least 
two different possibilities: a) A paramagnetic Fermi liquid, where 
the absence of magnetic ordering is just a consequence of the Pauli exclusion principle
(only a small fraction of the $f$-moment is screened by the conduction electrons) \cite{Bonca}; and
b) A partially polarized FM metal (see also \cite{nos1,nos2})
The existence of this second scenario as a possible solution of   
the PAM provides a natural explanation for a number of U (US, USe, UTe \cite{santini}, 
URu$_{2-x}$M$_{x}$Si$_{2}$ with M = Re, Tc and Mn \cite{Dalichaouch}) and Ce 
based (CeSi$_{x}$ \cite{Yashima}, CeGe$_2$ \cite{Lahiouel},
Ce(Rh$_{1-x}$Ru$_x$)$_3$B$_2$ \cite{Malik}, CeSi$_{1.76}$Cu$_{0.24}$ \cite{Boni}, 
Ce$_3$Bi$_4$, and CeNi$_{0.8}$Pt$_{0.2}$\cite{Fillion}) materials which exhibit the 
coexistence of ferromagnetism and  Kondo behavior. 

In addition,
we have established a connection between the strong hyridization limit of 
the PAM and the strong coupling limit of the KLM which is valid in any dimension.
This connection provides a physical meaning for the strong coupling 
regime of the KLM.

{\it Acknowledgements.} This work was sponsored by the US DOE. 
J. B. acknowledges the support Slovene Ministry of
Education Science and Sports and FERLIN.


\end{multicols}

\end{document}